\documentclass{article}
\usepackage{amssymb}

\usepackage{amsmath}

\input amssym.def
\input amssym
\newtheorem{proposition}{Proposition}
\newtheorem{theorem}{Theorem}

\newenvironment{proof}[1][
Proof]{\textbf{#1.} }{\ \rule{0.5em}{0.5em}}

\begin{document}

\title{The role of entanglement in quantum measurement and information processing}
\author{Paul Busch \\
%EndAName
{\small Department of Mathematics, University of Hull, Hull HU6 7RX}\\
{\small Electronic address: P.Busch@hull.ac.uk}}
\date{3 March 2003}
\maketitle

\begin{abstract}
The significance of the quantum feature of entanglement between
physical systems is investigated in the context of quantum
measurements. It is shown that, while there are measurement
couplings that leave the object and probe systems non-entangled,
no information transfer from object to probe can take place unless
there is there is at least some intermittent period where the two
systems are entangled.
\newline
PACS numbers: 03.65.Ca; 03.65.Ta; 03.65.Wj; 03.65.Ud.
\end{abstract}

\section{Introduction}

In recent years, the quantum feature of entanglement between
physical systems has increasingly  been recognized as an
enormously valuable resource for purposes of information
processing. Here we analyze the role of entanglement in the
context of measurement processes.

In contrast to the situation in classical physics where all
observables can be measured, in prinicple, with arbitrary accuracy
and with negligible disturbance, quantum measurements are subject
to the following theorem: there is no information gain without
\emph{some} state disturbance. The proof (Ref. \cite{QTM}, p. 32)
is simple: assuming that the state transformer (also known as
\emph{instrument} defined by a measurement scheme leaves
\emph{all} object states unchanged, it then follows immediately
that the probabilities of measurement outcomes do not depend on
the initial state of the measured object. This is to say that the
measured observable is \emph{trivial}, that is, represented by a
positive operator measure whose effects are multiples of the
identity operator of the underlying Hilbert space.

Hence there is no measurement with no disturbance. This
disturbance is caused by the interaction that takes place between
the object and the measurement device, usually mediated through a
probe system. But one usually thinks of interactions producing
entanglement. Hence the question arises as to whether measurements
are possible which leave the object and probe in a non-entangled
state, or whether there is a similar theorem to the above one that
says,``no measurement without entanglement". We will show that the
answer is affirmative, but not in the most direct sense
conceivable.

\section{Entanglement and Measurement}

We prove the following.

\begin{proposition}
Let $\mathcal{H}_{1},\mathcal{H}_{2}$ be complex separable Hilbert spaces, $%
\phi _{0}$ a unit vector in $\mathcal{H}_{2}$. Assume $U:\mathcal{H}%
_{1}\otimes \mathcal{H}_{2}\rightarrow \mathcal{H}_{1}\otimes \mathcal{H}%
_{2} $ is a unitary map such that for all $\varphi \in \mathcal{H}_{1}$, $%
U(\varphi \otimes \phi _{0})=\varphi ^{\prime }\otimes \phi ^{\prime }$ for
some unit vectors $\varphi ^{\prime }\in \mathcal{H}_{1}$, $\phi ^{\prime
}\in \mathcal{H}_{2}$. Then $U$ acts in one of the following two ways:%
\newline
(1) $U(\varphi \otimes \phi _{0})=V(\varphi )\otimes \phi ^{\prime }$, where
$V$ is an isometry in $\mathcal{H}_{1}$ and $\phi ^{\prime }$ is a fixed
unit vector in $\mathcal{H}_{2}$;\newline
(2) $U(\varphi \otimes \phi _{0})=\varphi ^{\prime }\otimes W_{12}\varphi $,
where $W_{12}$ is an isometry from $\mathcal{H}_{1}$ to $\mathcal{H}_{2}$
and $\varphi ^{\prime }$ is a fixed unit vector in $\mathcal{H}_{1}$.
\end{proposition}

\noindent \proof Let $\{\varphi _{n}:n=1,2,\dots \}$ be an orthonormal basis
of $\mathcal{H}_{1}$. There are systems of unit vectors $\varphi
_{n}^{\prime }\in \mathcal{H}_{1}$, $\phi _{n}^{\prime }\in \mathcal{H}_{2}$
such that $U\varphi _{n}\otimes \phi _{0}=\varphi _{n}^{\prime }\otimes \phi
_{n}^{\prime }$. Due to the unitarity of $U$, all the vectors $\varphi
_{n}^{\prime }\otimes \phi _{n}^{\prime }$ are mutually orthogonal. We show
that one of two cases (a), (b) must hold:\newline
(a) $\{\varphi _{n}^{\prime }\}_{n\in \mathbb{N}}$ is an orthonormal system,
all $\phi _{n}^{\prime }$ are parallel to $\phi _{1}^{\prime }$;\newline
(b) $\{\phi _{n}^{\prime }\}_{n\in \mathbb{N}}$ is an orthonormal system,
all $\varphi _{n}^{\prime }$ are parallel to $\varphi _{1}^{\prime }$.

For two vectors $\psi ,\xi $ which are mutually orthogonal, $\langle \psi
|\xi \rangle =0$, we will write $\psi \perp \xi $. If $\psi ,\xi $ are
parallel, we write $\psi \parallel \xi $. Since $U$ is unitary, this map
send orthogonal vector pairs to orthogonal pairs. Hence from $\varphi
_{1}\perp \varphi _{2}$ it follows that\ $\varphi _{1}^{\prime }\perp
\varphi _{2}^{\prime }$ or $\phi _{1}^{\prime }\perp \phi _{2}^{\prime }$.
Consider the first case. Then
\begin{equation}
U\left( \tfrac{1}{\surd 2}(\varphi _{1}+\varphi _{2})\otimes \phi
_{0}\right) =\varphi _{12}^{\prime }\otimes \phi _{12}^{\prime }=\tfrac{1}{%
\surd 2}\varphi _{1}^{\prime }\otimes \phi _{1}^{\prime }+\tfrac{1}{\surd 2}%
\varphi _{2}^{\prime }\otimes \phi _{2}^{\prime },
\end{equation}
where $\varphi _{12}^{\prime }\in \mathcal{H}_{1}$, $\phi _{12}^{\prime }\in
\mathcal{H}_{2}$ are some unit vectors. Since $\varphi _{1}^{\prime }\perp
\varphi _{2}^{\prime }$, it follows that $\phi _{2}^{\prime }=c\phi
_{1}^{\prime }$ with some $c\in \mathbb{C}$, $|c|=1$. Hence we have
\begin{equation}
U\left( (\varphi _{1}+\varphi _{2})\otimes \phi _{0}\right) =(\varphi
_{1}^{\prime }+c\varphi _{2}^{\prime })\otimes \phi _{1}^{\prime }
\end{equation}
Still considering the case $\varphi _{1}^{\prime }\perp \varphi _{2}^{\prime
}$, the relation $\varphi _{2}\perp \varphi _{3}$ implies that $\varphi
_{2}^{\prime }\perp \varphi _{3}^{\prime }$ or $\phi _{2}^{\prime }\perp
\phi _{3}^{\prime }$. Suppose the latter holds. We show that this leads to a
contradiction. Indeed this assumption gives $\varphi _{3}^{\prime
}=c^{\prime }\varphi _{2}^{\prime }$ and thus
\begin{eqnarray}
U\left( (\varphi _{1}+\varphi _{2}+\varphi _{3})\otimes \phi _{0}\right) &=&{%
\surd 3}\varphi _{123}^{\prime }\otimes \phi _{123}^{\prime } \\
&=&\varphi _{1}^{\prime }\otimes \phi _{1}^{\prime }+\varphi _{2}^{\prime
}\otimes \phi _{2}^{\prime }+\varphi _{3}^{\prime }\otimes \phi _{3}^{\prime
} \\
&=&(\varphi _{1}^{\prime }+c\varphi _{2}^{\prime })\otimes \phi _{1}^{\prime
}+\varphi _{3}^{\prime }\otimes \phi _{3}^{\prime }
\end{eqnarray}
where $\varphi _{123}^{\prime }$ and $\phi _{123}^{\prime }$ are some unit
vectors. Recalling that $\phi _{2}^{\prime }=c\phi _{1}^{\prime }$ and, by
assumption, $\phi _{2}^{\prime }\perp \phi _{3}^{\prime }$, then $\phi
_{1}^{\prime }\perp \phi _{3}^{\prime }$, and we see that $\varphi
_{1}^{\prime }+c\varphi _{2}^{\prime }=c^{\prime \prime }\varphi
_{3}^{\prime }$ for some $c^{\prime \prime }\neq 0$. Upon taking the inner
product of both sides with $\varphi _{1}^{\prime }$, we get (since $\varphi
_{1}^{\prime }\perp \varphi _{2}^{\prime }$) that $\langle \varphi
_{1}^{\prime }|\varphi _{1}^{\prime }\rangle =c^{\prime \prime }\langle
\varphi _{1}^{\prime }|\varphi _{3}^{\prime }\rangle =0$ (since $\varphi
_{3}^{\prime }=c^{\prime }\varphi _{2}^{\prime }\perp \varphi _{1}^{\prime }$%
). Hence $\varphi _{1}^{\prime }=0$ which is a contradiction.

Thus the assumption is false and we can only have $\varphi _{2}^{\prime
}\perp \varphi _{3}^{\prime }$. Continuing inductively, we obtain that $%
\{\varphi _{i}^{\prime }:i\in \mathbb{N}\}$ is an orthonormal system and all
$\phi _{i}^{\prime }=c_{i}\phi _{1}^{\prime }$. Therefore, we obtain
possibility (a) in the present case. Linearity then entails that $U(\varphi
\otimes \phi _{0})=V(\varphi )\otimes \phi _{0}^{\prime }$ for all $\varphi
\in \mathcal{H}_{1}$ and some isometric map $V$..

A completely analogous consideration can be applied in the second case of $%
\phi _{1}^{\prime }\perp \phi _{2}^{\prime }$, thus leading to the
possibility (b) and
\begin{eqnarray}
U(\varphi \otimes \phi _{0}) &=&\sum_{i}\langle \varphi _{i}|\varphi \rangle
U(\varphi _{i}\otimes \phi _{0})  \notag \\
&=&\varphi _{0}^{\prime }\otimes \sum_{i}\langle \varphi _{i}|\varphi
\rangle \phi _{i}^{\prime }=:\varphi _{0}^{\prime }\otimes W_{12}(\varphi )
\end{eqnarray}
for all $\varphi \in \mathcal{H}_{1}$ and some isometric map $W_{12}:%
\mathcal{H}_{1}\rightarrow \mathcal{H}_{2}$. $\square $

With this result we are ready to prove the following.

\begin{theorem}
Let $U:\mathcal{H}_{1}\otimes \mathcal{H}_{2}\rightarrow \mathcal{H}%
_{1}\otimes \mathcal{H}_{2}$ be a unitary mapping such that for all vectors $%
\varphi \in \mathcal{H}_{1}$, $\phi \in \mathcal{H}_{2}$, the image of $%
\mathcal{H}_{1}\otimes \mathcal{H}_{2}$ under $U$ is of the form $U(\varphi
\otimes \phi )=\varphi ^{\prime }\otimes \phi ^{\prime }$. Then $U$ is one
of the following:\newline
(a) $U=V\otimes W$ where $V:\mathcal{H}_{1}\rightarrow \mathcal{H}_{1}$ and $%
W:\mathcal{H}_{2}\rightarrow \mathcal{H}_{2}$ are unitary;\newline
(b) $U\left( \varphi \otimes \phi \right) =V_{21}\phi \otimes W_{12}\varphi $%
, where $V_{21}:\mathcal{H}_{2}\rightarrow \mathcal{H}_{1}$ and $W_{12}:%
\mathcal{H}_{1}\rightarrow \mathcal{H}_{2}$ are surjective isometries.%
\newline
The latter case can only occur if $\mathcal{H}_{1}$ and $\mathcal{H}_{2}$
are Hilbert spaces of equal dimensions.
\end{theorem}

\noindent \textbf{Proof.} Let $\{\phi _{i}:i=0,1,2,\dots \}$ be an
orthonormal basis of $\mathcal{H}_{2}$. For each $i$, we have, from
Proposition 1, that either $U(\varphi \otimes \phi _{i})=V_{i}(\varphi
)\otimes \phi _{i}^{\prime }$ ($\star $) with some unitary $V_{i}$, or $%
U(\varphi \otimes \phi _{i})=\varphi _{i}^{\prime }\otimes
W_{12}^{(i)}\varphi $ ($\star \star $) for some isometry $W_{12}^{(i)}$.

Case 1: $U(\varphi \otimes \phi _{0})=V_{0}\varphi \otimes \phi _{0}^{\prime
}$. We show that ($\star $) must hold for all $i$. Assume ($\star \star $)
holds for some $i\geq 1$. Considering the superposition $\phi _{0}+\phi _{i}$
we find, by an argument analogous to one used in the previous proof, that
either $V_{0}\varphi \perp \varphi _{i}^{\prime }$ and $\phi _{0}^{\prime
}\parallel W_{12}^{\left( i\right) }\varphi $, or $\phi _{0}^{\prime }\perp
W_{12}^{\left( i\right) }\varphi $ and $V_{0}\varphi \parallel \varphi
_{i}^{\prime }$. The first case would violate the isometric nature of $%
W_{12}^{\left( i\right) }$ and the second the unitarity of $V_{0}$. Hence ($%
\star \star $) is excluded and ($\star $) must hold for all $i$ in Case 1.

Still in Case 1, we therefore must have one of (A) $V_{0}\varphi \perp
V_{1}\varphi $ and $\phi _{0}^{\prime }\parallel \phi _{1}^{\prime }$; or
(B) $V_{0}\varphi \parallel V_{1}\varphi $ and $\phi _{0}^{\prime }\perp
\phi _{1}^{\prime }$ Consider case (A). Two possibilities arise (considering
the superposition $\phi _{0}+\phi _{2}$): either $V_{0}\varphi \perp
V_{2}\varphi $ and $\phi _{0}^{\prime }\parallel \phi _{2}^{\prime }$; or $%
V_{0}\varphi \parallel V_{2}\varphi $ and $\phi _{0}^{\prime }\perp \phi
_{2}^{\prime }$. The second leads to $\phi _{1}^{\prime }\perp \phi
_{2}^{\prime }$ (since $\phi _{0}^{\prime }\parallel \phi _{1}^{\prime }$),
therefore $V_{1}\varphi \parallel V_{2}\varphi $ (considering the
superposition $\phi _{1}+\phi _{2}$) and thus $V_{0}\varphi \parallel
V_{1}\varphi $, which contradicts the assumption of case (A). Hence in that
case one must always have $V_{0}\varphi \perp V_{i}\varphi $ and $\phi
_{0}^{\prime }\parallel \phi _{i}^{\prime }$ for all $i\geq 1$. Hence $\phi
_{i}^{\prime }=c_{i}\phi _{0}^{\prime }$, with $\left| c_{i}\right| =1$. But
that implies, for $\phi =\sum_{i}\alpha _{i}\phi _{i}$, that $U\varphi
\otimes \phi =\sum_{i}\alpha _{i}V_{i}\varphi \otimes \phi _{i}^{\prime
}=\left( \sum_{i}\alpha _{i}c_{i}V_{i}\varphi \right) \otimes \phi
_{0}^{\prime }$. This would contradict the surjectivity of $U$.

This leaves us with case (B). Suppose we have $\phi _{0}^{\prime }\parallel
\phi _{2}^{\prime }$ and thus $V_{0}\varphi \perp V_{2}\varphi $; this gives
$\phi _{2}^{\prime }\perp \phi _{1}^{\prime }$ \ (from $\phi _{0}^{\prime
}\perp \phi _{1}^{\prime }$) and so $V_{1}\varphi \parallel V_{2}\varphi $,
hence $V_{0}\varphi \perp V_{1}\varphi $ (from $V_{0}\varphi \perp
V_{2}\varphi $), in contradiction to (B). Therefore we must have $\phi
_{0}^{\prime }\perp \phi _{2}^{\prime }$ and by extension of this argument, $%
\phi _{0}^{\prime }\perp \phi _{i}^{\prime }$. Furthermore, since ($\star $)
holds in Case 1, similar arguments (considering superpositions $\phi
_{i}+\phi _{j}$, $\phi _{i}+\phi _{k}$, $\phi _{j}+\phi _{k}$) show that we
must have always $\phi _{i}^{\prime }\perp \phi _{j}^{\prime }$ and $%
V_{i}\varphi \parallel V_{j}\varphi $ for $i\neq j$. We thus obtain $%
V_{i}\varphi =c_{i}V_{0}\varphi $. It is not hard to see (considering $%
U\left( (\alpha \varphi +\beta \psi )\otimes \phi _{i}\right) $) that the
constants $c_{i}$ are independent of $\varphi $. We get $U(\varphi \otimes
\phi _{i})=V_{0}\varphi \otimes c_{i}\phi _{i}^{\prime }$. Unitarity of $U$
enforces that $V$ is unitary and the $\phi _{i}^{\prime }$ form an
orthonormal basis. Therefore we can define a unitary map $W$ as the unique
linear extension of $W\phi _{i}:=c_{i}\phi _{i}^{\prime }$. This finally
leads to $U\left( \varphi \otimes \phi \right) =V_{0}\varphi \otimes W\phi $.

Case 2: $U\left( \varphi \otimes \phi _{0}\right) =\varphi _{0}^{\prime
}\otimes W_{0}\varphi $. Suppose $U\left( \varphi \otimes \phi _{1}\right)
=V_{1}\varphi \otimes \phi _{1}^{\prime }$. This gives either $\varphi
_{0}^{\prime }\perp V_{1}\varphi $ and $W_{0}\varphi \parallel \phi
_{1}^{\prime }$, or $\varphi _{0}^{\prime }\parallel V_{1}\varphi $ and $%
W_{0}\varphi \perp \phi _{1}^{\prime }$. Both possibilities are excluded as $%
W_{0}$ and $V_{1}$ (being isometric maps) do not map onto a ray. We conclude
that in Case 2, $U\left( \varphi \otimes \phi _{i}\right) =\varphi
_{i}^{\prime }\otimes W_{i}\varphi $ must hold for all $i$.

Consider $U\left( \varphi \otimes \phi _{1}\right) =\varphi _{1}^{\prime
}\otimes W_{1}\varphi $. We must have either $\varphi _{1}^{\prime }\perp
\varphi _{0}^{\prime }$ and $W_{0}\varphi \parallel W_{1}\varphi $, or $%
\varphi _{1}^{\prime }\parallel \varphi _{0}^{\prime }$ and $W_{0}\varphi
\perp W_{1}\varphi $. In the latter case, suppose $\varphi _{2}^{\prime
}\perp \varphi _{0}^{\prime },$which goes along with $W_{0}\varphi \parallel
W_{2}\varphi $. This gives $\varphi _{1}^{\prime }\perp \varphi _{2}^{\prime
}$ and so $W_{1}\varphi \parallel W_{2}\varphi $, and therefore $%
W_{0}\varphi \parallel W_{1}\varphi $, in contradiction to the present case.
Therefore, if $\varphi _{1}^{\prime }\parallel \varphi _{0}^{\prime }$ then $%
\varphi _{i}^{\prime }\parallel \varphi _{0}^{\prime }$ for all $i\geq 1$.
As in Case 1, this violates the surjectivity of $U$.

Hence we must have the former case, $\varphi _{1}^{\prime }\perp \varphi
_{0}^{\prime }$ and $W_{0}\varphi \parallel W_{1}\varphi $. Again in analogy
to Case 1, we can conclude that $\varphi _{i}^{\prime }\perp \varphi
_{j}^{\prime }$ and $W_{i}\varphi \parallel W_{0}\varphi $ for all $i,j$. We
may write $W_{i}\varphi =c_{i}W_{0}\varphi $, where the $c_{i}$ are of
modulus 1 and independent of $\varphi $. Thus we get $U\left( \varphi
\otimes \sum_{i}\alpha _{i}\phi _{i}\right) =\sum_{i}\alpha _{i}c_{i}\varphi
_{i}^{\prime }\otimes W_{0}\varphi $. Putting $W_{12}:=W_{0}$, $\alpha
_{i}=\langle \phi _{i}|\phi \rangle $, and $V_{21}\phi
:=\sum_{i}c_{i}\langle \phi _{i}|\phi \rangle \varphi _{i}^{\prime }$, we
get the final result $U\left( \varphi \otimes \phi \right) =V_{21}\phi
\otimes W_{12}\varphi $. Again, unitarity of $U$ ensures that $W_{0}$ is
unitary and the $\varphi _{i}^{\prime }$ form an orthonormal basis, so that $%
V_{12}$ is also unitary. $\square $

A unitary map with the property that product states are sent to product
states can be used to model dynamics that do not lead to entanglement
between the systems involved. Thus it can be said that all non-entangling
dynamics are of the form described in the theorem above.

\textbf{Example. }Let $\mathcal{H}_{1}=\mathcal{H}_{2}=\mathcal{H}$. Let $%
E:\Sigma \rightarrow \mathcal{L}\left( \mathcal{H}\right) $ be a positive
operator valued measure (POVM) in $\mathcal{H}$, defined on a $\sigma $%
-algebra of subsets of some set $\Omega $, with values in the space of
bounded operators on $\mathcal{H}$. Define $U\left( \varphi \otimes \phi
\right) =\phi \otimes \varphi $. Then we have
\begin{equation}
\langle U\varphi \otimes \phi |I\otimes E\left( X\right) U\varphi \otimes
\phi \rangle =\langle \varphi |E\left( X\right) \varphi \rangle .
\end{equation}
This is the probability reproducibility condition which makes the
present model, with coupling $U$, pointer $E$, and initial probe
state $\phi $ a measurement scheme for the observable $E$ of the
first system \cite{QTM, OQP}.

This model demonstrates positively that information can be copied
from the object onto a probe in such a way that these two systems
are left non-entangled. Our theorems also show that there are two
distinct types of non-entangling unitary maps: product operators
or swap maps. Consider a continuous unitary group $U_t$ which
models the interaction between object and probe from time $t=0$ to
time $t=\tau$. Suppose $U_t$ is of the form $V_t\otimes W_t$ for
all $t$, $0\le t<\tau$. If $U_\tau$ were to have the form
$V_{21}\otimes W_{12}$, then continuity would dictate that, as
$t\to\tau$, then $V_t\varphi\to\phi$ for all $\varphi$, and
$W_t\phi\to\varphi$ for all $\varphi$. But this is clearly
impossible.

It follows that if a unitary continuous measurement dynamics $U_t$
leads to a state transformation $U_\tau$ given by the swap
mapping, then for $0<t<\tau$, some of the $U_t$ must be such that
they produce entanglement; they cannot all be of product form.

\section{Conclusion}

We conclude that abstract Hilbert space quantum mechanics admits
non-entangling measurements for all positive operator measures,
although intermediately some entanglement must build up. Whether
such measurement dynamics can be implemented by realistic
interactions is another question.

\end{document}